# A Machine Learning-Enhanced Benders Decomposition Approach to Solve the Transmission Expansion Planning Problem under Uncertainty

Stefan Borozan, Spyros Giannelos, Paola Falugi, Alexandre Moreira, Goran Strbac

*Abstract*—The necessary decarbonization efforts in energy sectors entail the integration of flexibility assets, as well as increased levels of uncertainty for the planning and operation of power systems. To cope with this in a cost-effective manner, transmission expansion planning (TEP) models need to incorporate progressively more details to represent potential long-term system developments and the operation of power grids with intermittent renewable generation. However, the increased modeling complexities of TEP exercises can easily lead to computationally intractable optimization problems. Currently, most techniques that address computational intractability alter the original problem, thus neglecting critical modeling aspects or affecting the structure of the optimal solution. In this paper, we propose an alternative approach to significantly alleviate the computational burden of large-scale TEP problems. Our approach integrates machine learning (ML) with the well-established Benders decomposition to manage the problem size while preserving solution quality. The proposed ML-enhanced Multicut Benders Decomposition algorithm improves computational efficiency by identifying effective and ineffective optimality cuts via supervised learning techniques. We illustrate the benefits of the proposed methodology by solving a number of multi-stage TEP problems of different sizes, based on the IEEE24 and IEEE118 test systems, while also considering energy storage investment options.

*Index Terms*--Benders decomposition, classification, expansion planning, machine learning, stochastic optimization, uncertainty

## Nomenclature

| | |
|---|---|
| CPI | Cut Performance Indicator |
| $N_{BD}$ | Maximum number of iterations in the Benders algorithm |
| $N_S$ | Number of sampling iterations |
| $N_\zeta$ | Number of classifiers |
| $\Theta_i^{CPI}$ | Training dataset for the corresponding CPI, $i = 1,..,N_\zeta$ |
| $\Omega_B$ | Set of all short-term representative blocks, indexed $b$ |
| $\Omega_C$ | Set of all Benders cuts, indexed $c$ |
| $\Omega_C^E$ | Set of all effective Benders cuts generated in iteration $k$ |
| $\Omega_{F(c)}^{CPI}$ | Set of values of $c$ for all features corresponding to CPI |
| $\Omega_M$ | Set of all scenario tree nodes, indexed $m$ |
| $\Omega_S$ | Set of all subproblems, indexed $s$ |
| $\Omega_\theta^{CPI}$ | Set of all labeling thresholds for the corresponding CPI |
| $\Omega_\zeta^{CPI}$ | Set of all classifiers of the corresponding CPI, indexed $\zeta$ |
| $\mathbb{P}^M$ | Master problem |
| $\mathbb{P}_s^S$ | Subproblem $s$ |
| $\epsilon_{BD}$ | Convergence tolerance of the Benders algorithm |
| $\epsilon_{PR}$ | Area under the Precision-Recall curve |
| $\epsilon_{ROC}$ | Area under the Receiver Operating Characteristic curve |
| $\pi_m$ | State probability of scenario tree node $m$ |
| $\bar{x}$ | Vector containing the current candidate solution of $\mathbb{P}^M$ |
| $\bar{\alpha}_s$ | $\mathbb{P}_s^S$ objective approximation at the current $\mathbb{P}^M$ solution |
| $\bar{\lambda}_s$ | Vector containing dual variables obtained from $\mathbb{P}_s^S$ |
| $y^\sigma$ | Slack decision variable representing load curtailment |
| $\bar{\psi}^M$ | Current objective function value of $\mathbb{P}^M$ |
| $\bar{\psi}_s^S$ | Current objective function value of $\mathbb{P}_s^S$ |

## I. Introduction

POWER system planning is a large-scale problem crucial in accommodating the cost-effective transition to a net-zero power sector. However, it is also subject to fundamental changes brought on by decarbonization efforts. Firstly, a major challenge is the significant multi-dimensional uncertainty around future system developments, as reflected in system pathways such as the Future Energy Scenarios in Great Britain [1]. Neglecting uncertainty altogether or even considering future scenarios deterministically risks underinvestment or overinvestment in network infrastructure – both associated with increased costs to society [2]. Therefore, it has been established in scientific literature that the application of a form of stochastic optimization method is essential for strategic decision-making [3], [4]. Furthermore, references [2] and [5] argue that multi-stage formulations allow anticipative strategic decision-making and are, therefore, necessary to unlock the potential for investment flexibility [6].

Further challenges are introduced with the importance of new technologies in response to the increased requirements for operational flexibility in low-carbon systems. Recent research has focused on modeling flexibility assets to both measure their impacts on expansion planning and to examine their roles as

Stefan Borozan, Spyros Giannelos, and Goran Strbac are with Imperial College London, London, United Kingdom.
Paola Falugi is with University of East London, London, United Kingdom.
Alexandre Moreira is with Lawrence Berkeley National Laboratory, Berkeley, CA, USA.
This work was supported by the Engineering and Physical Sciences Research Council (EPSRC) under Grant numbers EP/R045518/1 and EP/W034204/1.

non-network investment alternatives to network reinforcement, which is more exposed to the adverse effects of uncertainty due to the robust nature and long lead times of infrastructure projects. Authors in [7] investigate the impact of the electrification of transport on the distribution expansion planning problem considering the presence of distributed assets. References [2] and [8] propose alternative approaches to transmission expansion planning (TEP) with energy storage as a non-network alternative, while the same is done for demand response, soft-open points, and coordinated voltage control in [9]. In [10], vehicle-to-grid charger placements are co-optimized with line reinforcements in a multi-stage stochastic framework to allow planners to hedge against long-term uncertainty while reducing expected system operation costs.

Increasing modeling complexities through detailed uncertainty representation and more extensive investment portfolios, including assets with different construction, performance and cost characteristics, evolves this class of planning problems leading to more informed decisions [11], albeit at the risk of problem intractability [3]. In general, state-of-the-art expansion planning formulations are mixed-integer linear or nonlinear programs [4], which are NP-hard and can be extremely difficult to solve, while multi-stage stochastic formulations additionally suffer the curses of dimensionality.

To deal with computational difficulties associated with problem size, various decomposition and parallelization techniques are often applied in the solution methods of stochastic optimization problems [3], [12]. In [5], for instance, a detailed literature review is provided about decomposition methods used in power systems with a focus on the expansion planning problem. Benders decomposition [13], [14] has been highlighted as an attractive framework to address this class of problems as its structure allows for straightforward disaggregation of investment and operation constraints by taking the binary investment decisions as complicating variables [12]. Researchers have attempted to enhance various aspects of the performance of Benders decomposition [15]. For example, multicut formulations [16], which append one cut per subproblem (SP) in every iteration, are often used to improve convergence, while parallelization of the SPs execution can speed up solution times [15]. However, the critical issue in multicut Benders decomposition (MBD) is that all binary decision variables are aggregated in the master problem (MP), the size of which grows linearly over iterations, becoming increasingly difficult to solve. It has been reported that more than 90% of total solution time is dedicated to MPs [15] and it could even lead to intractability after a number of iterations. Only 3.13% of reviewed articles in [15] utilize a method directly aimed at controlling the size of the MP.

The interest in Machine Learning (ML) methods in power systems has been growing recently, as evidenced in [17] and [18]. The previous references reveal that ML has predominantly found applications in time-critical domains, where it can lead to significant improvements in execution times. Authors in [19] compare the application of reinforcement learning and convex optimization for the unit commitment problem. They argue the speed advantages of the ML-based approach, but also emphasize its limitations with respect to the suboptimal solutions that it produces. Therefore, the use of ML must be carefully considered before replacing proven optimization-based methods, especially in applications where solution time is not a critical characteristic.

Another way to exploit the strengths of ML is to integrate it within optimization frameworks, as argued in [20] and demonstrated with application to power systems in [21]. In line with the classification approach employed in [20], [22] identifies four categories of learning-assisted power system optimization algorithms: boundary parameter improvement, optimization option selection, surrogate model, and hybrid model. The review cites only two references with relevance to planning, highlighting the research gap. One example is [23], where the power flow calculation is replaced with a learning-based method in a hybrid model formulation. Other recent publications in this emerging field have focused on electric vehicle charging infrastructure planning in a predict-then-optimize formulation, such as [24] in which charging demand is first predicted using a graph convolutional network to inform the optimal charger allocation problem. The relevance of these works notwithstanding, none of them address the computational bottleneck of combinatorial optimization problems directly.

Promising techniques outlined in [20] are yet to be widely adopted in power system applications. In [25], an ML model is proposed to identify sets of active constraints with which optimal solutions can be obtained more efficiently. A similar method based on learning active constraints is proposed in [26] for linear bilevel problems with application to generator strategic bidding, and [27] proposes to learn redundant constraints from previous instances of the unit commitment problem to enhance the computational performance of mixed-integer programs. However, such methods change the problem structure and are heuristic, without guarantees of optimality or feasibility [26]. Novel approaches have recently emerged that are specifically aimed at classifying constraints in iterative solution algorithms. Notably, [28] proposes learning-based selection of cutting planes in integer programming methods, while [29] and [30] translate that idea to Benders decomposition of two-stage stochastic programs and of mixed-integer nonlinear programs in wireless communications, respectively. Nonetheless, to the authors' knowledge, this novel theory has neither been extended to a multi-stage stochastic setting nor been leveraged to address power systems problems to date.

Within this context, in this work we propose a computationally efficient methodology to address the multi-stage stochastic TEP problem. This methodology leverages ML techniques to enhance the solution algorithm, which is based on generating cutting planes to approximate future-cost functions. More specifically, we build upon previous research on cuts classification to propose a Benders decomposition approach suitable for large-scale MILP problems with complete recourse and able to manage the critical computational bottleneck – the explosion of MP size over iterations [3]. The proposed method is an ML-enhanced node-variable multicut Benders decomposition algorithm (ML-MBD) that can solve the multi-stage stochastic TEP problem under uncertainty, considering network and non-network investment options. We present an investigation into feature and target selection for this class of problems, as well as an appropriate sampling and training methodology. Finally, it is worth mentioning that this work presents a novel implementation of ML in power systems that tackles common tractability issues in large stochastic problems without resorting to reducing problem size through approaches that could jeopardize solution quality.

The contributions of this paper can be summarized as:

- Develops a novel computationally efficient ML-enhanced Benders decomposition method to solve the multi-stage stochastic TEP problem.
- Proposes a hybrid framework that exploits advantages of both ML and optimization for solving large multi-stage stochastic optimization problems with complete recourse and preserves the solution quality.
- Investigates and proposes suitable classification targets, features, and hyperparameters for this class of problems.
- Demonstrates the computational benefits and potential applications of the proposed method.

The paper begins with the mathematical formulations that form the basis for the developed ML-enhanced decomposition approach. Then, Section III presents the proposed method, including details on target selection, labeling, feature selection, and associated algorithms. The case studies that validate ML-MBD and demonstrate its benefits are presented in Section IV, while Section V provides a critical discussion on the developed method and its contributions. The conclusions are summarized in Section VI.

## II. MATHEMATICAL FORMULATIONS

We briefly introduce the multi-stage stochastic TEP problem and its decomposition, while a comprehensive discussion of stochastic programming can be found in [31].

Long-term uncertainty in stochastic optimization problems with a discrete uncertainty distribution can intuitively be represented using a scenario tree [31] in which each node in $\Omega_m$ is a realization of the uncertain parameters and associated with a state probability ($\pi_m$). On the other hand, short-term operational variability is considered in TEP through representative time blocks $\Omega_B$ that are typically not presented as part of the scenario tree but are directly associated with each $m$. The TEP under uncertainty problem can be compactly formulated as a node-variable multi-stage stochastic program with (1) – (6).

$$\psi = \min_{x_m, y_{m,b}} \sum_{m \in \Omega_m} \pi_m \left( f_m^I(x_m) + \sum_{b \in \Omega_B} f_{m,b}^O(y_{m,b}) \right) \quad (1)$$

subject to

$$h_m^I(x_m) = 0 \quad (2)$$
$$g_m^I(x_m) \leq 0 \quad (3)$$
$$h_m^O(x_{a(m)}, y_{m,b}) = 0 \quad (4)$$
$$g_m^O(x_{a(m)}, y_{m,b}) \leq 0 \quad (5)$$
$$x_m \in \mathcal{X}_m, \ y_{m,b} \geq 0, \ \forall m \in \Omega_m \quad (6)$$

The superscripts $I$ and $O$ relate to investment and system operation, respectively. The objective function minimizes the total expected cost across all considered realizations of uncertainty, comprising investment and system operation costs. The problem is a stochastic MILP because investment decisions $x_m$ are mixed integer and chosen from a portfolio of options, including network reinforcements and non-network alternatives, while decision variables $y_{m,b}$ are continuous and related to system operation. In (4) and (5) it is observed that system operation constraints are subject to investment decisions in all ancestor nodes of $m$, and including $m$, taking into consideration any construction lead times.

A problem is said to have complete recourse when for any feasible x, there exist y such that the problem is feasible in all stages. The stochastic TEP is a complete recourse problem because for all feasible $x_m$, system operation is feasible in all $m$ due to the presence of load curtailment as a slack decision variable ($y_{m,b}^\sigma$) in $f_{m,b}^O(\cdot)$, which is penalized with a large cost – the Value of Lost Load.

### A. Multicut Benders Decomposition

By taking the investment decisions as complicating variables, (1) – (6) can be decomposed into an investment problem ($\mathbb{P}^M$) and $|\Omega_s|$ system operation problems ($\mathbb{P}_s^S$), each corresponding to a unique $(m, b)$ pair. Letting $c_m, d_m, A_m, b_m, F_{m,b}$ and $h_{m,b}$ be functions of uncertainty, (1) is rewritten as:

$$\psi = \min_{x_m, y_{m,b}} \sum_{m \in \Omega_M} \pi_m \left( c_m^T x_m + \sum_{b \ in \ \Omega_B} d_m^T y_{m,b} \right) \quad (7)$$

Introducing the continuous variable $\alpha_{m,b} \in \mathbb{R}^+, \forall m, b$ to approximate operational costs, we formulate the $\mathbb{P}^M$ in iteration $k$ with (8) – (11).

$$\psi^{M(k)} = \min_{x_m, \alpha_{m,b}} \sum_{m \in \Omega_M} \pi_m \left( c_m^T x_m + \sum_{b \ in \ \Omega_B} \alpha_{m,b} \right) \quad (8)$$

subject to

$$A_m x_m \leq b_m, \ \forall m \in \Omega_M \quad (9)$$
$$\alpha_{m,b} \geq \left( h_{m,b} - F_{m,b} x_m \right)^T v_{m,b},$$
$$\forall m \in \Omega_M, b \in \Omega_B, v_{m,b} \in \Omega_{\mathcal{V}(m,b)}^{(k-1)} \quad (10)$$
$$x_m \in \mathcal{X}_m, \ \forall m \in \Omega_m \quad (11)$$

In every iteration, the MP yields a candidate solution $(\bar{x}_m, \bar{\alpha}_{m,b})$. The set $\Omega_{\mathcal{V}(m,b)}^{(k-1)}$ in (10) contains the identified extreme points of the feasible region of the $\mathbb{P}_s^S$ corresponding to $(m, b)$ at the current solution and is associated with the optimality cuts. Benders' method additionally involves feasibility cuts associated with the identified extreme rays of $\mathbb{P}_s^S$, which have been omitted in the presented formulation as all SPs are feasible for any MP solution, as previously explained. By fixing investment decisions to $\bar{x}_m$, $\mathbb{P}_s^S$ are free from any integer decision variables and defined for each $(m, b)$ pair as:

$$\psi_{m,b}^{S(k)}(\bar{x}_m) = \min_{y_{m,b}} d_m^T y_{m,b} \quad (12)$$

subject to

$$W_{m,b} y_{m,b} \leq h_{m,b} - F_{m,b} \bar{x}_m \ : \ \bar{\lambda}_{m,b} \quad (13)$$
$$y_{m,b} \geq 0 \quad (14)$$

The MP is a relaxation of the original problem, the objective function of which is reconstructed by iteratively generating optimality cuts of the form (10) from the solution of (12) – (14) and appending them in (8) – (11) until convergence is achieved. By weak duality, $\bar{\psi}^M$ provides a lower bound (LB) to the original problem, while a valid upper bound (UB) is given by $\sum_{m \in \Omega_M} \pi_m (c_m^T \bar{x}_m + \sum_{b \text{ in } \Omega_B} \bar{\psi}^S_{m,b}(\bar{x}_m))$, where $\bar{\psi}^M$ and $\bar{\psi}^S_{m,b}$ are the optimal values of the MP and each SP at the current candidate solution, respectively. Each appended cut increases computational burden equally, but its contribution to the rate of convergence is different. Therefore, the aim of ML-MBD is to discard those cuts that provide no or little contribution.

## III. Learning-enhanced Benders Decomposition method

In this section, we extend the novel theory in cuts classification to a multi-stage multicut formulation and propose appropriate revisions where previous research falls short for the application to power system optimization problems with complete recourse. The ML-MBD involves two main stages: 1) data sampling and classifiers training, and 2) problem execution. Three alternative versions of the method are developed. We first discuss the ML techniques used, the target selection, the labelling process, and the feature selection. We then provide a detailed description of the algorithms.

### A. ML method

We consider the use of supervised learning techniques to train the classifiers in this research. Compared to other options, they have the advantage of being relatively well understood in the field of artificial intelligence, with simpler implementation, fast convergence, and no requirement of a very large training dataset – which is crucial in our class of problems.

To maximize the performance of the proposed method, we explore and evaluate different ML techniques [32], as listed in Subsection III. C. Nonetheless, models based on Support Vector Machines (SVM), Decision Trees (DT), and Random Forests (RF) presented the best performance. As authors in [30] argue the importance of correctly classifying effective cuts, the precision-recall (PR) characteristics [32] of the ML models actually have a direct impact on the ML-MBD's performance. Models trained with RF incurred in better PR scores and are therefore used to train classifiers in the presented case studies.

### B. Target selection and labeling

The target variable for the ML models is in fact a cut performance indicator (CPI) that informs on the effectiveness of an individual cut in improving the approximation of the future cost function within the Benders algorithm. References [29] and [30] both propose the LB as target, which would be the natural choice given that it is the objective value of the MP and it increases monotonically. However, using the LB as CPI may lead to inefficient behavior of the UB and slower convergence as a result, an aspect that previous references do not investigate. Furthermore, the Benders UB does not behave predictably, so appending only the cuts that cause its value to decrease could be extremely valuable, thus worth exploring as a CPI. In the case studies, we identify shortcomings of both LB and UB as CPIs and additionally explore a third combined approach that mitigates limitations by applying the LB-based then the UB-based method sequentially. In summary, we propose and evaluate three different versions of the ML-MBD method:

1) ML-MBD-L (CPI based on LB)
2) ML-MBD-U (CPI based on UB)
3) ML-MBD-C (CPI based on both LB and UB)

We introduce a cut performance metric (CPM) that provides information on the improvement of the CPI as a result of appending a cut. As it is a continuous measure, a label transformation function is required to separate effective and ineffective cuts for the binary classification. Equations (15) and (16) define the CPM with reference to the CPI before and after cut $c$ has been added and evaluate it against a threshold $\theta$. Cuts that satisfy the inequalities are labeled 1, and 0 otherwise.

$$CPM_c^{LB} = \frac{LB^{(after)} - LB^{(before)}}{LB^{(after)}} \geq \theta^{LB} \quad (15)$$

$$CPM_c^{UB} = \frac{UB^{(after)} - UB^{(before)}}{UB^{(after)}} \leq \theta^{UB} \quad (16)$$

Unlike the monotonically increasing LB, changes in UB can be both positive and negative, so $\theta^{UB}$ is a negative value since improvement implies a decrease in the UB. The threshold is an important factor in ML-MBD as it ultimately influences the conservativeness of the classifier. A stricter $\theta^{CPI}$ would result in fewer cuts appended to the MP, risking slower convergence, while an overly permissive $\theta^{CPI}$ could render the use of ML redundant. The choice of threshold is challenging for two main reasons. Firstly, because there is no established rule as to what constitutes an effective cut [33]. And secondly, because the rate of convergence of BD has a tailing off characteristic such that the changes in the LB and UB diminish in latter iterations, causing distribution shift and implying that a single threshold might not be appropriate for all cuts. For these reasons, we use a set of thresholds with decreasing strictness, similar to the approach in [29]. We define $N_\zeta$ thresholds in $\Omega_\theta^{CPI} = \{\theta_1^{CPI}, \dots, \theta_{N_\zeta}^{CPI}\}$ for each CPI (where CPI is a placeholder for either LB or UB), based on observations of $CPM_c^{LB}$ and $CPM_c^{UB}$ for each cut $c$ in the respective training datasets. Thresholds in $\Omega_\theta^{LB}$ are chosen uniformly with values between, but excluding, the largest and the smallest observed $CPM_c^{LB}$, and the same is done for $\Omega_\theta^{UB}$ with values greater than the smallest observed $CPM_c^{UB}$ up to, and including, zero. This approach to thresholds selection was adopted because it generalizes well to any TEP problem.

### C. Feature selection

The aim of feature selection is to identify cut characteristics and complementary information that best inform the ML model on a cut's potential for CPI improvement. It is important to base this selection on both ML metrics and ML-MBD performance. We aim to define problem-independent features such that the method generalizes to a range of modified TEP problems, as discussed further in Subsection IV. B. 3) and Section V.

Here we investigate ten potential features, in combinations of two or more and with LB and UB as CPI. We evaluate their correlation and feature importance (Gini Importance and Mean

Decrease in Accuracy), as well as classification scores (receiver operating characteristic (ROC), balanced accuracy, and F-score) of models trained with SVM, DT, RF, logistic regression, and k-nearest neighbors [32]. The combinations with highest scores are selected to support ML-MBD-L and ML-MBD-U in solving two trial TEP problems, resulting in 72 test examples upon which the final feature selection is made based on the trade-off between number of cuts and number of iterations at convergence, as shown in Fig. 1 for all converging problems with an optimal cost within 1.5% of that obtained using MBD.

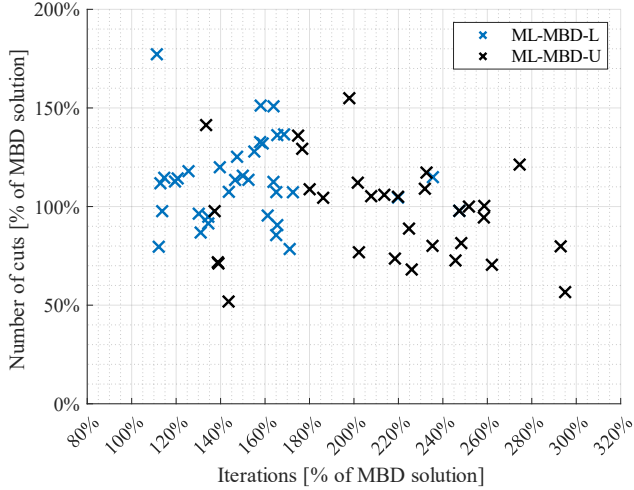

Fig. 1. Trade-off between number of cuts and number of iterations of the test models as a percentage of the non-ML MBD solution

Table I summarizes the investigated features and the observations in terms of their efficacy. Note that quantities here are defined on the set $\Omega_s$, where each $s$ translates to a unique $(m, b)$ in line with notation in Section II. The symbol $\Delta$ denotes the difference between the values in iterations $(k)$ and $(k-1)$.

The first feature is cut origin, namely the SP from which the cut originates. The second is the iteration in which the cut is generated, which might carry useful information given the tailing off effect of MBD. Then, Cut Violation is a feature proposed in [29] that is related to the feasible space of the MP that can be removed by appending the cut. It is inherently a useful measure, however, note that it is defined only at the current MP candidate solution $(\bar{x}, \bar{\alpha}_s)$ and resulting $\bar{\lambda}_s$, without any indication on how it translates to the rest of the feasible region and, by extension, to subsequent candidate solutions. Although this feature might contain useful information and is shown empirically to be an important feature, it does not carry full information and could be developed further in future work.

The objective function value of the SP where the cut is generated and the corresponding dual variables are contained within Cut Violation. Nonetheless, our investigation reveals that they could act as complementary features that increase the performance of ML-MBD. A similar observation is made for the final two features – solution proximity and value of the slack variable. The slack variable is penalized with a large cost in the SP objective function and could be considered as a reflection on the quality of the MP candidate solution. Intuitively, it should carry important information about the cuts in a problem with complete recourse.

TABLE I FEATURE SELECTION SUMMARY

| Feature | CPI = LB | | CPI = UB | |
|---|---|---|---|---|
| | Observation | Use | Observation | Use |
| Cut origin: $\mathbb{P}_s^S$ | detrimental | | irrelevant | |
| Iteration: $k$ | informative | ✓ | informative | ✓ |
| Cut Violation: $(h_s - F_s\bar{x})^T \bar{\lambda}_s - \bar{\alpha}_s$ | informative | ✓ | informative | ✓ |
| SP objective: $\bar{\psi}_s^S$ | irrelevant | | informative (supplementary) | ✓ |
| Change in SP objective: $\Delta\bar{\psi}_s^S$ | informative (supplementary) | ✓ | informative (supplementary) | |
| Dual variables: $\bar{\lambda}_s$ | irrelevant | | irrelevant | |
| Change in dual variables: $\Delta\bar{\lambda}_s$ | informative (supplementary) | ✓ | irrelevant | |
| Solution proximity: $|\Delta\bar{x}|$ | informative (supplementary) | ✓* | informative (supplementary) | ✓ |
| Value of slack variable: $\bar{y}_s^\sigma$ | informative (supplementary) | | informative (supplementary) | ✓ |
| Change in value of slack variable: $\Delta\bar{y}_s^\sigma$ | informative (supplementary) | ✓ | informative (supplementary) | |

There are a few observations to note on the final feature selection marked with (✓) in the table. Firstly, different features are preferred depending on the CPI. For instance, the performance of ML-MBD-L is improved when supporting features are formulated as a change in value from the previous iteration, while that is not the case for ML-MBD-U. Similarly, solution proximity is used for both CPIs, but the binary decision variables are excluded from ML-MBD-L features. Secondly, our investigation finds that the two-feature approach in [29] leads to underperformance of the ML-MBD, both in terms of ML metrics and number of cuts at convergence, and that the additional features proposed in [30] are not applicable to our decomposition technique. Lastly, unlike in both previous papers, which focus on other types of problems, we find that cut origin is not an effective predictor for a cut's effectiveness in our multi-stage formulation, and in fact it is detrimental to the performance of ML-MBD-L.

### D. Sampling and training procedures

The proposed offline sampling and training procedure is summarized in Algorithm 1 and consists of two parts – solving a TEP problem with a modified MBD algorithm to extract cut information and subsequently training the ML models. It covers the use of all or any of the three proposed versions of the method. Should only one CPI be used, then the actions related to the other can be neglected. Note that the sampling problem should be of similar structure as the one on which the ML-MBD will be applied, for example involving the same network and a simplified uncertainty representation.

In the first part, the MDB is modified so as to reveal unique information on the effectiveness of each cut individually. This is achieved by re-solving the MP with only one newly appended cut and measuring the improvement in the LB as a direct consequence of that cut. Then, all cuts are appended to progress the algorithm and continue with SPs execution. The improvement in UB is measured at the end of the iteration as an aggregated contribution of all cuts appended in that iteration. To obtain individualized UB improvement information, similar to LB, $|\Omega_S|^2$ additional SP instances would need to be solved, which prohibitively increases the sampling time for problems with large scenario trees and the benefit of this information does not justify the increased time demand for smaller problems. Finally, the features are extracted at the end of each iteration.

---
**Algorithm 1** Data sampling and classifiers training
**Step 0: Initialization**
Set $N_\zeta, \epsilon_{ROC}^{min}, \epsilon_{PR}^{min}, \Omega_c^{(0)} = \emptyset, \Theta_i^{LB} \leftarrow \emptyset, \Theta_i^{UB} \leftarrow \emptyset$
Set $k = 0, UB^{(0)} = \infty, LB^{(0)} = 0, \epsilon_{BD}, N_S, j = 0$
**Step 1: Sample using modified MBD**
$j = j + 1$
**While** $k \leq j \times N_S$ and $\frac{UB^{(k)}-LB^{(k)}}{LB^{(k)}} > \epsilon_{BD}$
    $k = k + 1$
    **For** all $c$ in $\Omega_c^{(k-1)}$
        Solve $\mathbb{P}^{M(k)}$ with $c$ and all cuts in $\Omega_c^{(0)} ... \Omega_c^{(k-2)}$, and obtain $\bar{\psi}_c^{M(k)}$
        **If** $k > 1$, compute $CPM_c^{LB} = \frac{|\bar{\psi}_c^{M(k)} - \bar{\psi}_c^{M(k-1)}|}{\bar{\psi}_c^{M(k)}}$
    Solve $\mathbb{P}^{M(k)}$ with all $c$ in $\Omega_c^{(0),...,(k-1)}$ and obtain $\bar{\psi}^{M(k)}$ and $\bar{x}^{(k)}$
    Solve $\mathbb{P}_s^{S(k)}$ $\forall s \in \Omega_s$, obtain $\bar{\psi}_s^{S(k)}$ and $\bar{\lambda}_s^{(k)}$, and save cuts in $\Omega_c^{(k)}$
    Extract features and save in $\Omega_{F(c)}^{LB}$ and $\Omega_{F(c)}^{UB}$, respectively, $\forall c \in \Omega_c^{(k)}$
    Compute $UB^{(k)}$ and $LB^{(k)}$
    **If** $k > 1$, compute $CPM_c^{UB} = \frac{|UB^{(k)}-UB^{(k-1)}|}{UB^{(k)}}$ $\forall c \in \Omega_c^{(k-1)}$
**Step 2: Labeling**
Determine $\Omega_\theta^{LB}$ based on all observed $CPM_c^{LB}$ values
**For** $i = 1,..,N_\zeta$ and $c = 1,...,|\Omega_c|$
    **If** $CPM_c^{LB} \geq \theta_i^{LB}, \Theta_i^{LB} \leftarrow \{\Omega_{F(c)}^{LB}, 1\}$
    **Else**              $\Theta_i^{LB} \leftarrow \{\Omega_{F(c)}^{LB}, 0\}$
Determine $\Omega_\theta^{UB}$ based on all $CPM_c^{UB}$ values
**For** $i = 1,..,N_\zeta$ and $c = 1,...,|\Omega_c|$
    **If** $CPM_c^{UB} \leq \theta_i^{UB}, \Theta_i^{UB} \leftarrow \{\Omega_{F(c)}^{UB}, 1\}$
    **Else**              $\Theta_i^{UB} \leftarrow \{\Omega_{F(c)}^{UB}, 0\}$
**Step 3: Training**
Perform under-sampling in $\Theta_i^{LB}$ and $\Theta_i^{UB}$ $\forall i \in \{1,..,N_\zeta\}$
Train $\zeta_1, \zeta_{median}$ and $\zeta_{N_\zeta}$ and obtain $\epsilon_{ROC}$ and $\epsilon_{PR}$ for each
**If** $\epsilon_{ROC} < \epsilon_{ROC}^{min}$ or $\epsilon_{PR} < \epsilon_{PR}^{min}$, for $\zeta_1, \zeta_{median}$ or $\zeta_{N_\zeta}$, go to **Step 1**
**Else** Finish training remaining models in $\Omega_\zeta$
**END**

---

Step 1 is run for $N_S$ iterations, such that a wide enough range of Benders cuts are sampled but overfitting of the ML models is evaded and is significantly fewer than the number of iterations required to achieve convergence. Then, $\Omega_\theta^{CPI}$ are determined for both LB and UB, and cut samples are labelled accordingly. With that, $N_\zeta$ datasets are created for each CPI ($\Theta_i^{LB}$ and $\Theta_i^{UB}$ in Algorithm 1) with $N$ rows and $|\Omega_F^{CPI}|+1$ columns. They contain the same number of rows (samples) and identical $|\Omega_F^{CPI}|$ columns (features), but the labels in the final column differ due to the varying strictness of the thresholds. Finally, an under-sampling procedure is performed for each dataset to balance the classes before training.

Classifiers $\zeta_1^{CPI}, \zeta_{med}^{CPI}$ and $\zeta_{N_\zeta}^{CPI}$ are trained on $\Theta_1^{CPI}, \Theta_{med}^{CPI}$ and $\Theta_{N_\zeta}^{CPI}$, respectively, where $\Theta_{med}^{CPI}$ is the dataset corresponding to the median threshold in $\Omega_\theta^{CPI}$. They are evaluated using their ROC and PR curves, such that if the areas under the ROC and PR curves are below pre-defined minimum levels ($\epsilon_{ROC}^{min}, \epsilon_{PR}^{min}$), the algorithm returns to Step 1 where the modified MBD is resumed for another $N_S$ iterations. The termination criterion is based on the performance of three ML models because each is trained with a different number of samples due to the class imbalance of the respective dataset and subsequent under-sampling, which could result in different performances of the models. Finally, the remaining classifiers in $\Omega_\zeta$ are trained. Note that different ML techniques can be used for training, as mentioned in Subsection III. A., and evaluated such that the final models used in the problem execution stage are the ones with best performance.

### E. Solution algorithm

The proposed solution method is summarized in Algorithm 2. Step 1 is performed only if suitable classifiers for the problem do not already exist, as supported by the generalization properties presented in the case studies.

In Step 2, cuts are classified using $\Omega_\zeta^{CPI} = [\zeta_1^{CPI},...,\zeta_{N_\zeta}^{CPI}]$, starting with the strictest model $\zeta_1^{CPI}$ until at least one cut is identified as effective. Should no cut be deemed effective by any classifier in $\Omega_\zeta^{CPI}$, then the algorithm proceeds as traditional MBD in that iteration. Stricter classifiers may become redundant considering the distribution shift as the Benders algorithm progresses and could therefore be discarded. However, we find that although CPI improvements decrease on average over iterations, they do not decrease monotonically and often a stricter classifier can still be utilized after it has failed to label any cuts as useful in a previous iteration, potentially yielding greater benefits compared to the approach in [29]. Finally, note that the presented algorithm covers the use of any of the three ML-MBD versions. For ML-MBD-C, the CPI is updated when the moving average of $CPM^{UB}$ in the previous 10 iterations ($\delta_{UB}$) falls below a pre-determined value $\epsilon_{UB}$. This action can be ignored in ML-MBD-U, while ML-MBD-L only requires the CPI to be adequately initialized at the start.

---
**Algorithm 2** ML-MBD
**Step 1: Offline training**
**If** $\Omega_\zeta = \emptyset$, execute **Algorithm 1**
**Step 2: Problem execution**
Set $k = 0, UB^{(0)} = \infty, LB^{(0)} = 0, N_{BD}, \epsilon_{BD}, \Omega_c^{(0)} = \emptyset, CPI = UB, \epsilon_{UB}$
**While** $k \leq N_{BD}$ and $\frac{UB^{(k)}-LB^{(k)}}{LB^{(k)}} > \epsilon_{BD}$
    $k = k + 1$
    **For** $i = 1,...,N_\zeta$
        Classify all $c \in \Omega_c^{(k-1)}$ using $\zeta_i^{CPI}$: If '1', $\{c\} \to \Omega_c^{E,(k-1)}$
        **If** $\Omega_c^{E,(k-1)} \neq \emptyset$, **break**
    **If** $\Omega_c^{E,(k-1)} = \emptyset$, $\{c\} \to \Omega_c^{E,(k-1)}$, $\forall c \in \Omega_c^{(k-1)}$
    **If** $|\delta_{UB}| < \epsilon_{UB}$, $CPI = LB$
    Solve $\mathbb{P}^{M(k)}$ with all $c$ in $\Omega_c^{E,(0),...(k-1)}$ and obtain $\bar{\psi}^{M(k)}$ and $\bar{x}^{(k)}$
    Solve $\mathbb{P}_s^{S(k)}$ $\forall s \in \Omega_s$, obtain $\bar{\psi}_s^{S(k)}, \bar{y}_s^{(k)}$, and $\bar{\lambda}_s^{(k)}$, and save cuts in $\Omega_c^{(k)}$
    Compute $UB^{(k)}$ and $LB^{(k)}$
Obtain optimal solution
**END**

---

### F. Computational effort

ML integration adds a certain overhead to the full solution process. The ML techniques employed require a relatively low number of samples to produce well-performing models and training time is negligible, especially considering the extensive solution time of large combinatorial optimization problems. In our investigations and case studies, the training time of a single classifier typically ranged between 0.5 and 3 seconds, with only a few instances taking between 20 and 30 seconds, depending on the number of samples and ML technique. The classification step in Algorithm 2 took at most 0.6 seconds, including time required for memory communication between scripts.

The bulk of the overhead is in fact in Step 1 of Algorithm 1 and it depends on the problem size and number of iterations required for generating an adequate dataset. In the studies presented in this paper, we do not impose any expectations or restrictions on the total training time and $N_S = 20$ is used throughout for consistency. In most cases, Algorithm 1 terminated after two or three repetitions of Step 1 for a high standard for $\epsilon_{ROC}^{min}$ and $\epsilon_{PR}^{min}$ of 0.92. Nevertheless, the sampling and training procedure does not add to the computational efforts of the problem execution itself because it is performed offline and only once for application to different problems, enabled by the generalization properties demonstrated in the case studies.

## IV. CASE STUDIES

In this section, we apply ML-MBD to the problem of co-expansion of the transmission system with battery energy storage (BES) as a non-network investment alternative. The computational benefits of ML-MBD are analyzed with respect to the number of appended cuts, which is directly related to the memory requirements of the problem, and the MP solution times. The results validate the proposed method and demonstrate its applicability in terms of scalability and generalizability. Optimization problems were implemented in FICO XPress 8.13 and the ML scripts in Python 3.9. Results were obtained on a Dual Xeon computer with 512 GB RAM. All cases were solved with $N_\zeta = 10$, $\epsilon_{BD} = 0.01$, and $N_{BD} = 1{,}000$.

### A. Description

The case studies are based on the IEEE 24-Bus Reliability Test System (IEEE24) and the IEEE 118-Bus System (IEEE118), the topologies of which can be found in [34]. Line capacities in both networks are modified to allow unconstrained operation in the initial system state, as explained in [5], while the IEEE24 case additionally includes solar and wind generation as done in [10]. Generator capacities are scaled up to ensure system adequacy throughout the planning horizon.

Three test cases of different sizes were developed subject to uncertainty in the level and rate of peak demand increase, as presented with the scenario trees in Fig. 2. All cases involve four decision-making stages over a 40-year horizon, but the number of scenarios and scenario tree nodes is different. The first tree shows 6 scenarios over 13 nodes, the second 18 scenarios and 27 nodes, and the final case considers 27 scenarios over 40 nodes. Fig. 2 displays scenario probabilities as well as peak load increase in each node with reference to the initial system state (root node). Each scenario tree node represents an investment decision-making point and is associated with a unique system operation problem. Short-term variability is accounted for with four representative blocks of system operation. Demand profiles are obtained from IEEE24 and IEEE118 data and renewable generation time-series from [10]. Since each $(m, b)$ pair represents a separate SP, the first case involves $|\Omega_S| = 52$, the second $|\Omega_S| = 108$, and the third $|\Omega_S| = 160$, and the same number of Benders cuts per iteration.

Line reinforcements are subject to an annualized fixed cost of \$121,600 per km, annualized variable cost of \$76 per MW and km, and a commissioning delay of one stage. BES units are assumed to have a discharging duration of 2 hours and 90 % efficiency. They are subject to an annualized investment cost of \$102,000 per MW and are available in the same stage in which the investment decision is made. Case studies on the IEEE24 involve the possibility to invest in BES in all buses, while 13 candidate buses are selected in IEEE118 in keeping with [5].

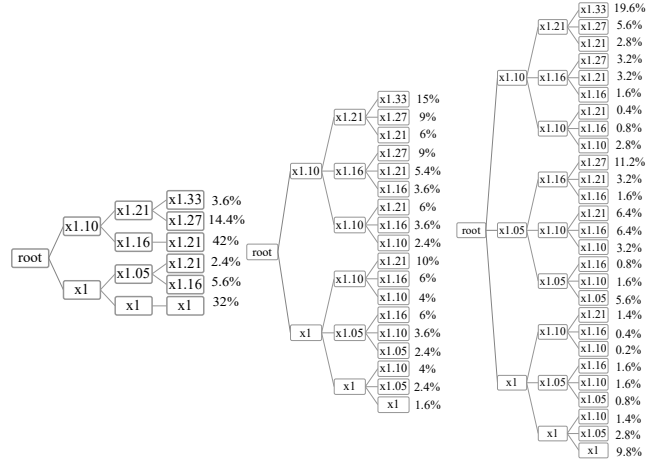

Fig. 2. Scenario trees describing the three test cases

### B. Results

#### 1) Validation and scalability

Table II summarizes the results at termination of the solution algorithm. Note that both cases with 40-node scenario trees solved using ML-MBD-U reached the maximum iteration limit before convergence. Moreover, we emphasize that the IEEE118 40-node problem failed to converge with MBD as it stopped progressing after 7 days and was terminated after 10 days.

As the contributions of this research concern the solution approach, we focus on the obtained costs and convergence to validate the method, while the precise investment strategies on the test problems are not of consequence. We stress, however, that all convergent ML-MBD cases result in identical investment decisions as with MBD. In terms of cost, it can be seen in the results table that all expected total system costs obtained using a version of the ML-MBD method are very close to the non-ML MBD solutions. Namely, most fall within 0.5% of the corresponding MBD reference values, except on two occasions with ML-MBD-U when costs differ by 0.86 % and 1.34 %. This signals that ML-MBD converges to the correct TEP solution, as evidenced by the values for best UB in Fig. 3.

TABLE II
RESULTS OF THE CASE STUDIES

| Network | Scenario tree | Method | Iterations | Total cuts | Solution gap | Total solution time [hours] | Best LB [£ million] | Best UB [£ million] |
|---|---|---|---|---|---|---|---|---|
| **IEEE24** | 13 nodes | MBD | 88 | 4,524 | 0.89% | 1.81 | 6,285 | 6,317 |
| | | ML-MBD-L | 127 | 4,885 | 0.91% | 2.64 | 6,261 | 6,314 |
| | | ML-MBD-U | 287 | 3,080 | 0.36% | 3.31 | 6,311 | 6,371 |
| | | ML-MBD-C | 116 | 4,398 | 0.86% | 2.16 | 6,268 | 6,314 |
| | 27 nodes | MBD | 60 | 6,372 | 0.58% | 9.57 | 21,964 | 22,175 |
| | | ML-MBD-L | 75 | 6,534 | 0.84% | 10.89 | 21,821 | 22,128 |
| | | ML-MBD-U | 164 | 2,110 | 0.91% | 13.01 | 21,715 | 22,227 |
| | | ML-MBD-C | 65 | 4,765 | 0.86% | 9.14 | 21,929 | 22,121 |
| | 40 nodes | MBD | 58 | 9,120 | 0.83% | 11.35 | 38,144 | 39,038 |
| | | ML-MBD-L | 67 | 7,569 | 0.77% | 10.52 | 38,180 | 38,504 |
| | | ML-MBD-U | 1,000 | 4,880 | 15.04% | 23.08 | 34,131 | 39,532 |
| | | ML-MBD-C | 63 | 5,970 | 0.81% | 9.78 | 38,146 | 38,516 |
| **IEEE118** | 13 nodes | MBD | 156 | 8,060 | 0.50% | 7.22 | 58,451 | 58,969 |
| | | ML-MBD-L | 175 | 6,434 | 0.84% | 6.23 | 57,542 | 58,460 |
| | | ML-MBD-U | 224 | 4,179 | 0.95% | 6.78 | 57,746 | 58,668 |
| | | ML-MBD-C | 179 | 6,276 | 0.72% | 6.35 | 57,538 | 58,476 |
| | 27 nodes | MBD | 250 | 26,892 | 0.96% | 35.44 | 56,353 | 56,679 |
| | | ML-MBD-L | 305 | 22,189 | 0.78% | 33.99 | 56,022 | 56,499 |
| | | ML-MBD-U | 812 | 13,319 | 0.93% | 59.00 | 55,891 | 56,403 |
| | | ML-MBD-C | 278 | 21,261 | 0.88% | 32.22 | 56,350 | 56,632 |
| | 40 nodes | MBD | 175 | 27,840 | 2.35% | 240 | 55,503 | 55,964 |
| | | ML-MBD-L | 232 | 26,542 | 0.85% | 165.15 | 55,519 | 55,946 |
| | | ML-MBD-U | 1,000 | 9,267 | 13.66% | 222.63 | 52,243 | 56,798 |
| | | ML-MBD-C | 226 | 25,702 | 0.96% | 159.90 | 55,503 | 55,957 |

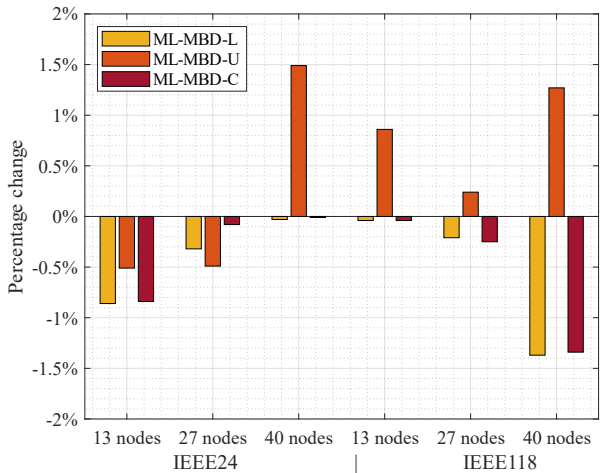

Fig. 3. Percentage change between the best upper bound obtained using ML-MBD methods and the reference values obtained using MBD

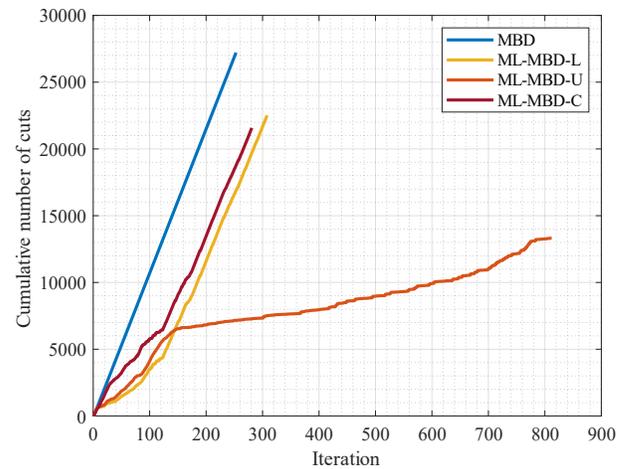

Fig. 4. Cumulative number of appended cuts over iterations for the four MBD methods in the IEEE118 27 nodes case

A general observation from Table II is that ML-MBD requires more iterations compared to MBD, but it consistently converges with a lower number of cuts. Using the IEEE118 and 27-node tree case as an example, Fig. 4 demonstrates the rate at which the MP grows using the four considered decomposition methods. MBD increases the problem size linearly, in this case adding 108 constraints every iteration. Conversely, all ML-MBD versions manage the size growth by discarding cuts that have been classified as ineffective, yet ultimately reaching an equivalent solution. This example also shows that the problem grows most conservatively with ML-MBD-U, converging with only 13,319 cuts, albeit at the expense of increased total solution time after 812 iterations.

To highlight the computational benefit of ML-MBD, Fig. 5 shows the percentage reduction in the total number of appended cuts, compared to the MBD solution as reference.

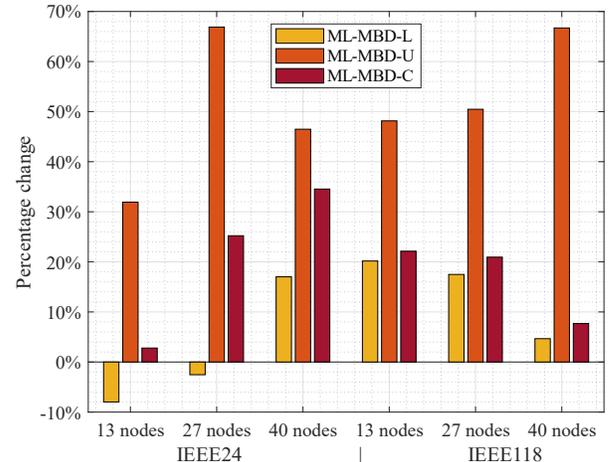

Fig. 5. Percentage reduction in the total number of cuts at convergence using ML-MBD methods compared to MBD

As previously illustrated, the problem grows the least with ML-MBD-U, which appends between 32% and 67% fewer cuts than MBD, but convergence within the limit is not always reached. ML-MBD-L underperforms in the smallest considered problems, but its benefits are evident in the larger problems and the algorithm converges every time. Finally, ML-MBD-C sits in between the other two versions in terms of MP size at convergence, while it also converges in each case. Note that for the largest problem, IEEE118 with 40-node scenario tree, the effect of all ML-MBD versions is in fact much more significant than the figure suggests because MBD fails to converge. As such, the percentage difference shown in this case can be regarded as the gap to the tractability limit. Taking ML-MBD-C as example, considering the difference of 2,138 cuts to MBD in its final completed iteration, ML-MBD-C would have been able to progress for at least 13 further iterations if all cuts were appended, or more if some were classified as ineffective and thus discarded.

*2) Computational performance*

The MP is NP-hard, meaning that it is exponential-time solvable and that the worst-case complexity increases exponentially with the problem size, but there are no proven rules regarding the average computational complexity of NP-hard problems. Furthermore, solution times in practice depend on additional factors, such as CPU usage. Therefore, the computational performance of the proposed method must be evaluated empirically and we analyze the results only for direct comparison between the ML and non-ML MBD on separate test cases. Notwithstanding, the results show a clear trend that less time is spent solving MPs with ML-MBD. Fig. 6 demonstrates the reduction in average MP solution time for ML-MBD-C compared to MBD. The improvements in computation time increase with the size of the TEP problem, up to 72.1 %. Note again that the final case does not reflect the full benefit of the proposed method because of the intractability issues for MBD, which failed to progress past iteration 175. The percentage reduction in MP solution time up to this iteration was 52.4 %.

The larger number of required iterations for ML-MBD may lead to an increase in total solution time, as previously noted for ML-MBD-U and observed in Table II for the two smallest problems. Note, however, that the impact of the number of iterations on total solution time could be largely reduced with parallel execution of SPs, which we avoid in order to identify direct effects of the application of ML-MBD. Nonetheless, the considerable MP solution time savings achieved with ML-MBD in the larger problems compensate for the time demands of additional iterations and lead to reductions in total solution times of up to 33.4 %, observed in the last case. This fact highlights the pertinence of ML-MBD to large-scale problems, especially those that challenge tractability limits.

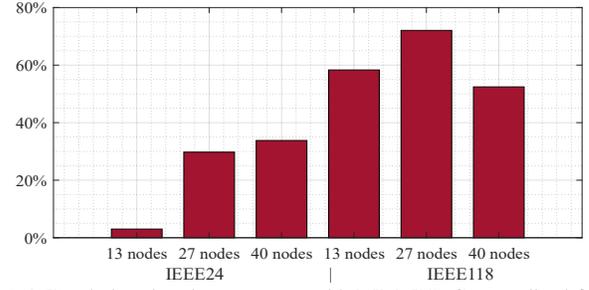

Fig. 6. MP solution time improvement with ML-MBD-C normalized for the number of iterations and as a percentage of the MBD reference values

*3) Generalizability*

We test the method's generalizability on modified problems with respect to aspects that might occur in implementation, such as changes to the assumptions on long-term uncertain parameters, short-term operational variability, scenario tree transition probabilities, or the size and shape of the scenario tree. Specifically, we take ML-MBD-C trained on the IEEE118 problem with a 13-node scenario tree and apply it directly, without re-training, to the following modified problems:

- IEEE118 and 13-node scenario tree with different transition probabilities
- IEEE118 and 13-node scenario tree with different long-term uncertainty assumptions on demand increase
- IEEE118 and 13-node scenario tree with different short-term variability time-series (representative blocks)
- IEEE118 and 27-node scenario tree

In all cases, ML-MBD-C converges to an equivalent solution to the one obtained with MBD, but with reduced computational burden. Fig. 7 shows the rate of MP size growth over iterations for the four cases and demonstrates that the proposed method is able to manage the problem size, despite the ML models being trained on a different problem. ML-MBD-C achieves a decrease in the total number of constraints in the MP of the final iteration of 18.9 %, 27.0 %, 15.5 % and 1.2 %, for the four cases respectively. Moreover, Fig. 8 highlights the significant MP solution time savings obtained. The results demonstrate that the proposed method generalizes exceptionally well to modified problems without repeated execution of Algorithm 1.

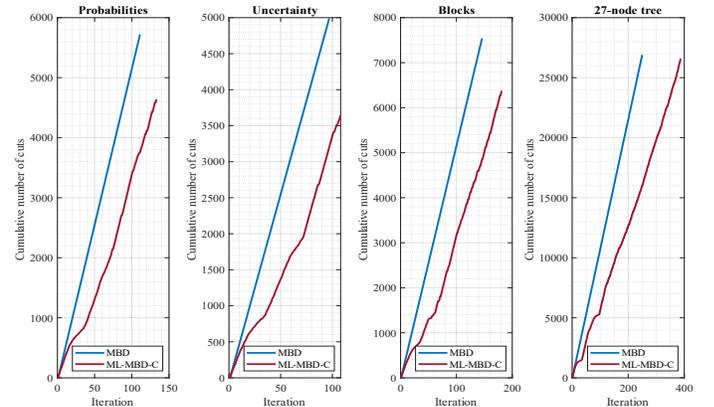

Fig. 7. Comparison between MBD and ML-MBD-C in terms of cumulative number of appended cuts over iterations for the four modified problems

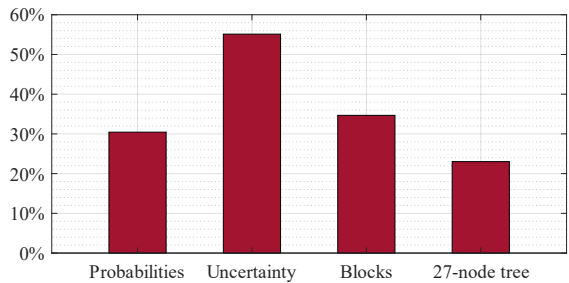

Fig. 8. MP solution time improvements with ML-MBD-C normalized for the number of iterations and as a percentage of the MBD reference values for the four modified problems

## V. Discussion

Modern TEP exercises are witnessing a trend of increasing complexity, leading to computationally intractable optimization problems. As such, modelling real-world details that allow for more informed decision-making is limited by computational scalability, which is one of the principal challenges in power system problems [3]. Techniques like scenario reduction or recent learning-based solutions have been applied to mitigate tractability issues, but such approaches alter the problem structure and may lead to different or suboptimal solutions. In contrast, the method developed in this paper preserves the original problem and leads to a solution of the same quality. Most notably, we demonstrate its application to solve a highly complex TEP problem that is intractable with a state-of-the-art decomposition method. The results in Section IV. additionally show that the ML-MBD is highly beneficial even for instances that can be solved with the non-ML-enhanced decomposition, as evidenced in Fig. 5 that shows the significant reduction in number of constraints and Fig. 6 that highlights the valuable MP solution time reduction. These observations are important because, as previously argued, the overwhelming bulk of the computational effort of MBD is in solving MP instances.

Three versions of the ML-MBD method are presented because each is applicable depending on user preference. As observed in Table II, ML-MBD-L is highly reliable, converging with a modest increase in number of iterations and with a lower best UB. However, while it leads to significant size reductions in the larger problems, it underperforms compared to MBD for the two smallest problems considered. ML-MBD-U, on the other hand, appends cuts at the most conservative rate, leading to convergence with the smallest number of MP constraints or failure to converge within 1,000 iterations, which may cause concern regarding its reliability. Finally, ML-MBD-C performs excellently across all previous considerations and would therefore be the preferred all-round method. It should be noted too, that all ML-MBD methods have exceptional ML performance scores.

Notwithstanding the low computational overhead of the ML-MBD, the sampling procedure involves solving a number of iterations of MBD. This fact is justified for two reasons: Firstly, ML-MBD could enable the convergence of certain problems that are otherwise intractable; and secondly, TEP studies typically involve solving the problem multiple times with modified inputs. In Subsection IV. B. 3), we show that the method generalizes to modified problems and it is this property that would allow the user to solve all subsequent studies without the need to re-sample and re-train ML models, thus increasing the marginal benefit of ML-MBD over MBD with every run.

ML-MBD finds its application to TEP under uncertainty for the aforementioned reasons. Furthermore, we note that the problem-independent feature selection employed could result in its applicability to problems with similar structure that require repeated execution, such as stochastic unit commitment, which could benefit from ML-MBD with faster MP computation or the opportunity to increase modelling complexity. Lastly, a limit of the proposed method is that it works to filter ineffective cuts and, as such, convergence cannot be achieved in fewer iterations than with MBD. Methods aimed at accelerating convergence through advanced cut generation exist, for instance as in [35]. Applying the ideas developed in this work to such techniques could unlock further computational benefits.

## VI. Conclusion

It is critical that modern TEP frameworks involve a high degree of modelling complexity that enables more informed decision-making to facilitate the transition to a net-zero power sector. However, state-of-the-art models are reaching computational limits, even with the application of advanced solution techniques. Motivated by the main computational bottleneck of MBD and the potential of learning-assisted optimization, this paper proposed an ML-enhanced Benders decomposition approach to solve large-scale stochastic TEP problems, which effectively manages the increase in MP size over iterations. The importance of the contribution has been corroborated by the attained results which, most notably, demonstrated that the ML-MBD is able to reach convergence when applied on a problem that is otherwise unsolvable with the well-established MBD approach, thus pushing the tractability limits of modern power system problems. Significant benefits were observed in comparably smaller problems as well, particularly in reducing computational time and memory requirements. We also emphasized the advantage of the proposed method in preserving solution quality, in contrast to other approaches that alter the original problem structure. Finally, the demonstrated generalizability properties highlight the value and applicability of ML-MBD in practice. In future work, we plan to implement the proposed decomposition in other power system problems and to improve the convergence rate of ML-MBD through advanced cut generation techniques.

## VII. References

[1] National Grid ESO, "Future Energy Scenarios 2021," July, 2021.
[2] I. Konstantelos and G. Strbac, "Valuation of Flexible Transmission Investment Options Under Uncertainty," IEEE Transactions on Power Systems, vol. 30, (2), pp. 1047-1055, 2015. DOI: 10.1109/TPWRS.2014.2363364.
[3] L.A. Roald, D. Pozo, A. Papavasiliou, D.K. Molzahn, J. Kazempour and A. Conejo, "Power systems optimization under uncertainty: A review of methods and applications," Electric Power Systems Research, vol. 214, 2023, DOI: 10.1016/j.epsr.2022.108725.
[4] A. J. Conejo et al, *Investment in Electricity Generation and Transmission: Decision Making Under Uncertainty*. Cham, Switzerland: Springer International Publishing, 2016.

[5] P. Falugi, I. Konstantelos and G. Strbac, "Planning With Multiple Transmission and Storage Investment Options Under Uncertainty: A Nested Decomposition Approach," IEEE Transactions on Power Systems, vol. 33, (4), pp. 3559-3572, 2018. DOI: 10.1109/TPWRS.2017.2774367

[6] A. K. Dixit and R. Pindyck, *Investment Under Uncertainty*. Princeton; Chichester: Princeton University Press, 1994.

[7] P. M. de Quevedo, G. Muñoz-Delgado and J. Contreras, "Impact of Electric Vehicles on the Expansion Planning of Distribution Systems Considering Renewable Energy, Storage, and Charging Stations," IEEE Transactions on Smart Grid, vol. 10, (1), pp. 794-804, 2019. DOI: 10.1109/TSG.2017.2752303.

[8] S. Dehghan and N. Amjady, "Robust transmission and energy storage expansion planning in wind farm-integrated power systems considering transmission switching," IEEE Transactions on Sustainable Energy, vol. 7, pp. 765–774, 2016.

[9] I. Konstantelos, S. Giannelos and G. Strbac, "Strategic Valuation of Smart Grid Technology Options in Distribution Networks," IEEE Transactions on Power Systems, vol. 32, (2), pp. 1293-1303, 2017. DOI: 10.1109/TPWRS.2016.2587999.

[10] S. Borozan, S. Giannelos, G. Strbac, "Strategic Network Expansion Planning with Electric Vehicle Smart Charging Concepts as Investment Options," Advances in Applied Energy, February 2022. DOI: 10.1016/j.adapen.2021.100077.

[11] S. Li and D. Tirupati, "Dynamic capacity expansion problem with multiple products: Technology selection and timing of capacity additions," Operations Research, vol. 42, pp. 958–976, 1994.

[12] A.J. Conejo, E. Castillo, R. Minguez, R. Garcia-Bertrand, *Decomposition Techniques in Mathematical Programming: Engineering and Science Applications*, Springer Science & Business Media, 2006.

[13] J. Benders, "Partitioning methods for solving mixed variables programming problems," Numerische Mathematik, vol. 4, 1962.

[14] R.M. Van Slyke and R Wets, "L-Shaped Linear Programs with Applications to Optimal Control and Stochastic Programming," SIAM Journal on Applied Mathematics, vol. 17, (4), pp. 638–63, 1969, http://www.jstor.org/stable/2099310.

[15] R. Rahmaniani, T.G. Crainic, M. Gendreau and W. Rei, "The Benders decomposition algorithm: A literature review," European Journal of Operational Research, vol. 259, (3), pp. 801-817, 2017, DOI: 10.1016/j.ejor.2016.12.005.

[16] J. R. Birge and F. V. Louveaux, "A multicut algorithm for two-stage stochastic linear programs," Eur. J. Oper. Res., vol.34, pp. 384–392, 1988.

[17] M. S. Ibrahim, W. Dong, and Q. Yang, "Machine learning driven smart electric power systems: Current trends and new perspectives," Applied Energy, vol. 272, 2020, DOI: 10.1016/j.apenergy.2020.115237.

[18] X. Chen, G. Qu, Y. Tang, S. Low and N. Li, "Reinforcement Learning for Selective Key Applications in Power Systems: Recent Advances and Future Challenges," in IEEE Transactions on Smart Grid, vol. 13, no. 4, pp. 2935-2958, July 2022, DOI: 10.1109/TSG.2022.3154718.

[19] C. O'Malley, P. de Mars, L. Badesa, G. Strbac, "Reinforcement Learning and Mixed-Integer Programming for Power Plant Scheduling in Low Carbon Systems: Comparison and Hybridisation," arXiv:2212.04824 [eess.SY], Dec. 2022, https://doi.org/10.48550/arXiv.2212.04824.

[20] Y. Bengio, A. Lodi, and A. Prouvost, "Machine learning for combinatorial optimization: A methodological tour d'horizon," European Journal of Operational Research, vol. 290, no. 2, pp. 405–421, 2021, DOI: 10.1016/j.ejor.2020.07.063.

[21] S. Pineda and J. M. Morales, "Is learning for the unit commitment problem a low-hanging fruit?," Electric Power Systems Research, vol. 207, June 2022, DOI: 10.1016/j.epsr.2022.107851.

[22] G. Ruan, H. Zhong, G. Zhang, Y. He, X. Wang and T. Pu, "Review of learning-assisted power system optimization," CSEE Journal of Power and Energy Systems, vol.7, 2021, doi: 10.17775/CSEEJPES.2020.03070.

[23] Y. Tao, J. Qiu and S. Lai, "A Supervised-Learning Assisted Computation Method for Power System Planning," in IEEE Transactions on Artificial Intelligence, vol. 3, no. 5, pp. 819-832, Oct. 2022, DOI: 10.1109/TAI.2021.3133821.

[24] C. Li, Z. Dong, G. Chen, B. Zhou, J. Zhang and X. Yu, "Data-Driven Planning of Electric Vehicle Charging Infrastructure: A Case Study of Sydney, Australia," in IEEE Transactions on Smart Grid, vol. 12, no. 4, pp. 3289-3304, July 2021, DOI: 10.1109/TSG.2021.3054763.

[25] S. Misra, L. Roald, and Y. Ng, "Learning for Constrained Optimization: Identifying Optimal Active Constraint Sets," INFORMS Journal on Computing, vol. 34, pp. 463–480, 2021, DOI: 10.1287/ijoc.2020.1037.

[26] E. Prat and S. Chatzivasileiadis, "Learning Active Constraints to Efficiently Solve Linear Bilevel Problems: Application to the Generator Strategic Bidding Problem," in IEEE Transactions on Power Systems, 2022, DOI: 10.1109/TPWRS.2022.3188432.

[27] A. S. Xavier, F. Qiu, and S. Ahmed, "Learning to Solve Large-Scale Security-Constrained Unit Commitment Problems," INFORMS Journal on Computing, vol. 33, pp. 739–756, 2021, DOI: 10.1287/ijoc.2020.1037.

[28] Y. Tang, S. Agrawal and Y. Faenza, "Reinforcement Learning for Integer Programming: Learning to Cut," in *Proceedings of the 37th International Conference on Machine Learning*, pp. 9367—9376, 2020, Available: https://proceedings.mlr.press/v119/tang20a.html.

[29] H. Jia and S. Shen, "Benders Cut Classification via Support Vector Machines for Solving Two-Stage Stochastic Programs," INFORMS Journal on Optimization, vol. 3, no. 3, pp. 278-297, Mar 2021, DOI: 10.1287/ijoo.2019.0050.

[30] M. Lee, N. Ma, G. Yu and H. Dai, "Accelerating Generalized Benders Decomposition for Wireless Resource Allocation," in IEEE Transactions on Wireless Communications, vol. 20, no. 2, pp. 1233-1247, Feb. 2021, DOI: 10.1109/TWC.2020.3031920.

[31] J. R. Birge and F. Louveaux, *Introduction to Stochastic Programming*, Springer New York, 2011, DOI: 10.1007/978-1-4614-0237-4.

[32] A.C. Müller and S. Guido, *Introduction to machine learning with Python: a guide for data scientists*. 1st ed. Sebastopol, CA, O'Reilly Media, 2017.

[33] A. Ruszczyński, "Decomposition Methods," Handbooks in Operations Research and Management Science, vol. 10, pp. 141–211, 2003, DOI: 10.1016/S0927-0507(03)10003-5.

[34] S. Peyghami, P. Davari, M. Fotuhi-Firuzabad and F. Blaabjerg, "Standard Test Systems for Modern Power System Analysis: An Overview," in IEEE Industrial Electronics Magazine, vol. 13, no. 4, pp. 86-105, Dec. 2019, DOI: 10.1109/MIE.2019.2942376.

[35] R. Brandenberg and P. Stursberg, "Refined cut selection for benders decomposition: applied to network capacity expansion problems," Math Meth Oper Res 94, 383–412, 2021, DOI: 10.1007/s00186-021-00756-8.